\documentclass[prd,showpacs]{revtex4}
\usepackage{amsmath}
\usepackage{amssymb}
\usepackage{latexsym}
\usepackage{enumerate}
\usepackage{bbm}
\usepackage{amsthm}
\usepackage{hyperref}
\usepackage{graphicx}
\usepackage{dcolumn}
\usepackage{bm}
\newcommand{\pound}{\emph{\textsterling}}
\newcommand{\be}{\begin{equation}}
\newcommand{\ee}{\end{equation}}
\newcommand{\ben}{\begin{equation*}}
\newcommand{\een}{\end{equation*}}
\newcommand{\bea}{\begin{eqnarray}}
\newcommand{\eea}{\end{eqnarray}}
\newcommand{\bean}{\begin{eqnarray*}}
\newcommand{\eean}{\end{eqnarray*}}

\newcommand{\bsub}{\begin{subequations}}
\newcommand{\esub}{\end{subequations}}

\newcommand{\disfrac}[1][2]{\displaystyle\frac}

\newcommand{\ima}{\mathbbmtt{i}}

\newcommand{\bbar}{\overline}

\begin{document}
\title{{\bf Minisuperspace Canonical Quantization of the Reissner-Nordstr\"om Black Hole via Conditional Symmetries}}
\vspace{1cm}
\author{T. Christodoulakis} \email{tchris@phys.uoa.gr} \author{N. Dimakis}\email{nsdimakis@gmail.com}
\author{Petros A. Terzis} \email{pterzis@phys.uoa.gr}
\affiliation{Nuclear and Particle Physics Section, Physics Department, University of Athens, GR 157--71 Athens}
\author{Babak Vakili}\email{b-vakili@iauc.ac.ir} \affiliation{Department of Physics, Chalous Branch, IAU, P.O. Box 46615-397, it Chalous, Iran}
\author{E. Melas}\email{evangelosmelas@yahoo.co.uk}
\affiliation{Logistics Department, GR32-200, Thiva, Technological Educational, Institution of Chalkida,
Electrical Engineering Department, GR 35--100, Lamia }
\author{Th. Grammenos} \email{thgramme@civ.uth.gr} \affiliation{Department of Civil Engineering, University of Thessaly, GR 383--34 Volos}

\begin{abstract}
We use the conditional symmetry approach to study the $r$-evolution
of a minisuperspace spherically symmetric model both at the
classical and quantum level. After integration of the coordinates
$t$, $\theta$ and $\phi$ in the gravitational plus electromagnetic
action the configuration space dependent dynamical variables turn
out to correspond to the $r$-dependent metric functions and  the
electrostatic field. In the context of the formalism for constrained
systems (Dirac - Bergmann, ADM) with respect to the radial
coordinate $r$, we set up a point-like reparameterization invariant
Lagrangian. It is seen that, in the constant potential
parametrization of the lapse, the corresponding minisuperspace is a
Lorentzian three-dimensional flat manifold which obviously admits
six Killing vector fields plus a homothetic one. The weakly
vanishing $r$-Hamiltonian guarantees that the phase space quantities
associated to the six Killing fields are linear holonomic integrals
of motion. The homothetic field provides one more rheonomic integral
of motion. These seven integrals are shown to comprise the entire
classical solution space, i.e. the space-time of a
Reissner-Nordstr\"om black hole, the $r$-reparametrization
invariance since one dependent variable remains unfixed, and the two
quadratic relations satisfied by the integration constants. We then
quantize the model using the quantum analogues of the classical
conditional symmetries, and show that the existence of such
symmetries yields solutions to the Wheeler-DeWitt equation which, as
a semiclassical analysis shows, exhibit a good correlation with the
classical regime. We use the resulting wave functions to investigate
the possibility of removing the classical singularities.

\vspace{5mm}\noindent
\end{abstract}


\maketitle

\numberwithin{equation}{section}


\pagebreak
\section{Introduction}
Symmetry considerations have acquired a very prominent role in all
branches of theoretical physics. This is probably due to the fact
that all conservation laws in physics are the result of some kind of
symmetry in the corresponding physical system. In this sense a
symmetry is a kind of variation of the Lagrangian of a dynamical
system that leaves the equations of motion invariant. One of the
most important types of such symmetries which has lots of
applications in classical mechanics and quantum field theory is the
well-known Noether symmetry. Mathematically, the famous Noether
theorem states that a vector field $X$ is a symmetry for a given
dynamical system if the Lie derivative of its Lagrangian along this
vector field vanishes ${\cal L}_XL=0$ (\cite{Noether},
\cite{Struck}). The first application, to the best of our knowledge,
of this criterion in constrained systems is given in \cite{Cap}.
Under this condition the vector field $X$ generates the conserved
currents from which the integrals of motion can be obtained (see
\cite{Cap1} - \cite{Darabi} for the applications of the Noether
symmetry approach in various cosmological models and black hole
physics). More generally, the symmetries of a Riemannian space may
also be represented by a vector field $X$ which satisfies an
equation of the form ${\cal L}_X {\bf A}={\bf B}$, where ${\bf A}$
and ${\bf B}$ are some geometric objects \cite{Tsa}. For instance,
in a Riemannian space with metric ${\cal G}_{\mu \nu}$, $X$ is a
conformal Killing vector if ${\bf A}={\cal G}_{\mu \nu}$ and ${\bf
B}=\phi(x^{\alpha}){\cal G}_{\mu\nu}$. In the case where
$\phi(x^{\alpha})=0$ the vector $X$ is known as a Killing vector and
when $\phi(x^{\alpha})$ is a non-vanishing constant $X$ is a
homothetic vector. There are also other kinds of such symmetries
that we will not mention here but a classification of them can be
found in \cite{Kat}.

In the canonical formulation of general relativity the space of all
Riemannian 3-dimensional metrics and matter fields on the spatial
hypersurfaces form an infinite-dimensional space, the so-called
superspace, which is the basic configuration space of quantum
gravity. However, in cosmology due to the many symmetries of the
underlying cosmological models the infinite degrees of freedom of
the corresponding superspace are truncated to a finite number and
thus a particular minisuperspace model is achieved. It is easy to
show that the evolution of such a system, when the equations of
motion are obtained from an action principle, can be produced by a
Lagrangian of the following form:

\begin{equation}\label{Int1}
L=\frac{1}{2n}G_{\alpha
\beta}(q)\dot{q}^{\alpha}\dot{q}^{\beta}-nV(q),
\end{equation}
where $q^{\alpha}$ and $n$ are the dependent dynamical variables and
the lapse function representing the coordinates of the
minisuperspace with metric $G_{\alpha \beta}(q)$, $V(q)$ is a
potential function and an overdot indicates derivation with respect
to some independent dynamical parameter. Since the dynamics of the
system in this formalism resembles the motion of a point particle
with coordinates $q^{\alpha}$ in a Riemannian space with metric
$G_{\alpha \beta}$, many interesting features may occur when this
space has some symmetries. In particular, one can define a
conditional symmetry generated by a vector field $\xi$ which is a
simultaneous conformal Killing vector field of the metric $G_{\alpha
\beta}(q)$ and the potential function $V(q)$, that is \cite{tchris1}
\begin{equation}\label{Int2}
{\cal L}_{\xi}G_{\alpha \beta}=\phi(q)G_{\alpha \beta},\hspace{5mm}{\cal L}_{\xi}V(q)=\phi(q)V(q).
\end{equation}
As noted above, each symmetry corresponds to a phase-space quantity
representing an integral of motion. In \cite{tchris1}, it is shown
that the integrals of motion resulting from (\ref{Int2}) can be
written as
\begin{equation}\label{Int3}
Q_I=\xi^{\alpha}_I p_{\alpha},
\end{equation}
where $p_{\alpha}=\frac{\partial L}{\partial \dot{q}^{\alpha}}$ is
the momentum conjugate to $q^{\alpha}$. In order to pass to the
quantum theory associated with these models, one should note that
the variation of (\ref{Int1}) with respect to $n$ yields
\begin{equation}\label{Int4}
\frac{1}{2n^2}G_{\alpha\beta}\dot{q}^{\alpha}\dot{q}^{\beta}+V(q)=0
\end{equation}
which, being the zero-energy condition, leads to the Hamiltonian
constraint
\begin{equation}\label{Int5}
H=n\left[\frac{1}{2}G^{\alpha \beta}p_{\alpha}p_{\beta}+V(q)\right]=n{\cal H}=0.
\end{equation}
Therefore, following the canonical quantization method, this
Hamiltonian gives rise to the Wheeler-DeWitt (WDW) equation
$\widehat{{\cal H}}\Psi(q)=0$, where $\Psi(q)$ is the wave function
of the quantized system and $\widehat{{\cal H}}$ should be written
in a suitable operator form. Now, it is easy to see that the Poisson
brackets of (\ref{Int3}) with the Hamiltonian vanish weakly on the
constrained surface. In the lapse parametrization $n=\frac{N}{V}$,
were the potential is constant, the aforementioned Poisson brackets
vanish identically. The quantum counterpart of this statement is
that the operator forms of (\ref{Int3}) and the scaled Hamiltonian
commute with each other which means that $\widehat{Q}_I$ and
$\widehat{{\cal H}}$ have simultaneous eigenfunctions. In summary,
the quantum counterpart of the theory with the aforesaid symmetry
can be described by the following equations (more details are
presented in the following sections):
\begin{eqnarray}\label{Int6}
\left\{
\begin{array}{ll}
\widehat{{\cal H}}\Psi(q)=0,\\
\widehat{Q}_I \Psi(q)=\kappa_I \Psi(q),
\end{array}
\right.
\end{eqnarray}
where $\kappa_I$ are the eigenvalues of $Q_I$.

In this paper we study the behavior of a static, spherically
symmetric space-time in the framework of the presence of conditional
symmetries in minisuperspace constrained systems. The phase-space
variables turn out to correspond to the $r$-dependent metric
functions and to an electrostatic field with which the action of the
model is augmented.

In section 2 we follow \cite{tchris1} - \cite{Cav2} and construct a
minisuperspace Lagrangian, in the form of (\ref{Int1}), using the
canonical decomposition along the radial coordinate $r$ which now
plays the role of a dynamical variable. We then deal with some
considerations on this minisuperspace constrained system possessing
conditional symmetries and by passing to the Hamiltonian formalism
we reveal six conditional symmetries and a rheonomic integral of
motion. Under these conditions we show that the classical solution
of such a system can be identified with the space-time of a
Reissner-Nordstr\"om (RN) black hole (\cite{Reis}, \cite{Nord}. For
higher dimensions see \cite{Nieto}).

In section 3 we consider the quantization of the system in which we
adopt the quantum analogues of the linear integrals of motion as
supplementary conditions imposed on the wave function, the latter
also satisfying, of course, the Wheeler-DeWitt quantum constraint.
To see how we can recover the classical solutions from the quantum
wave functions, we present a semiclassical analysis of the model
above described in section 4. The curious and interesting situation
of the vanishing quantum potential is investigated and fully
explained in section 5. Finally, some concluding remarks are
included in the discussion.

\section{Classical formulation and conditional symmetries} \label{sect2}
The  general form of a static, spherically symmetric line element is
\be \label{metric} ds^2=-a^2(r)dt^2+n^2(r)dr^2+b^2(r) \left(
d\theta^2 +  \sin^2\theta\, d\phi^2 \right), \ee with $n(r)$ playing
the role of the $r$-lapse function, while $a(r)$ and $b(r)$ are the
``dynamical"  dependent variables in the $r$-foliation. In order to
acquire the RN solution we need to consider an electrostatic field
minimally coupled to gravity. Thus, the full action is written as
\be \label{actionpr} S_{g+em} =\int \! \mathcal{L}_{GR}\, d^4 x
+\int \! \mathcal{L}_{EM}\, d^4 x =\int \! \sqrt{-g}\,R\, d^4 x -
\int\! \sqrt{-g}\, F_{\mu\nu}F^{\mu\nu} \, d^4 x, \ee where
$F_{\mu\nu}=A_{\mu , \nu}-A_{\nu , \mu}$ is the antisymmetric
electromagnetic tensor and $A_{\mu}$ is the potential with
$A_0=f(r)$ and $A_{1}=A_{2}=A_{3}=0$. In \eqref{actionpr} we have
chosen the units $c=1$, $G=\frac{1}{4\pi}$. The variation of this
action with respect to the space-time metric $g_{\mu\nu}$ leads to
Einstein's field equations \be \label{einequ} E_{\mu\nu}=2\,
T_{\mu\nu}, \ee where $E_{\mu\nu}=R_{\mu\nu}-\frac{1}{2}\, R\,
g_{\mu\nu}$ is the Einstein tensor and $T_{\mu\nu}=
F_{\mu\kappa}F_{\nu}^{\phantom{\nu}\kappa}-\frac{1}{4}\,
g_{\mu\nu}F_{\kappa\lambda}F^{\kappa\lambda}$ is the stress-energy
tensor associated with the electromagnetic field.

The variation with respect to the field $A_\mu$ leads to the
equations of motion \be \label{max} F^{\mu\nu}_{\phantom{\mu\nu}
;\mu}=0, \ee which together with the consistency conditions (being
in fact identities due to the assumed form of $F_{\mu\nu}$) \be
\label{well} F_{\mu\nu;\kappa}+ F_{\kappa\mu;\nu}+F_{\nu\kappa;\mu}
\equiv 0, \ee form the complete set of Maxwell's equations. One can
check that, using line element \eqref{metric}, equations \eqref{max}
are satisfied whenever equations \eqref{einequ} hold. Thus, for the
determination of the classical solution space it suffices to solve
the latter.

Apart from the field theory approach, one can be led to effectively
the same equations of motion by integration of the redundant degrees
of freedom in action \eqref{actionpr}, i.e. integrating over $t$,
$\theta$ and $\phi$ and ignoring a multiplicative (infinite)
constant. All system information is then contained in a reduced,
point-like action $S=\int\! L(a,b,f,a',b',f',n)\, dr$ with the
following Lagrange function:
 \be \label{Lagv} L=\frac{1}{2\, n}
\left(8 \,b \, a'\,b' + 4\, a\, b'^2+4 \, \frac{b^2}{a} f'^2\right)
+ 2\, n \, a, \ee where $'$ denotes differentiation with respect to
the spatial coordinate $r$. It is easy to verify that the
Euler-Lagrange equations ensuing from \eqref{Lagv} are equivalent to
the reduced Einstein's equations obtained by the substitution of the
line element \eqref{metric} and $A_\mu=(f(r),0,0,0)$ in
\eqref{einequ}. The Lagrangian \eqref{Lagv} belongs to a particular
form of singular Lagrangians: $L=\frac{1}{2\, n}\, G_{\mu\nu} \,
{q'}^\mu\,{q'}^\nu + n\, V(q)$. If one uses the freedom to
reparametrize the lapse, then \eqref{Lagv} can be brought to a form
in which the potential $V$ is constant. In our case we choose to set
$n= \frac{N}{2\, a}$, which leads to \be \label{Lag} L=\frac{1}{2\,
N} \left(16\, a \,b \, a'\,b' + 8\, a^2\, b'^2+ 8 \, b^2 f'^2\right)
+ N, \ee or, in a more concise form, $L=\frac{1}{2\, N}\,
\bbar{G}_{\mu\nu} \, {q'}^\mu\,{q'}^\nu + N$ with ${q'}^\mu = (a',
b', f')$ and \be \label{supmet} \bbar{G}_{\mu\nu} =
\begin{pmatrix}
0 & 8\, a\, b & 0 \\ \\
8\, a\, b & 8\, a^2 & 0 \\ \\
0 & 0 & 8 b^2
\end{pmatrix} .
\ee

As shown in \cite{tchris1}, it is in this particular lapse
parametrization that the conditional symmetries of the phase-space,
as defined in \cite{Kuchar}, become Killing vector fields of the
supermetric \eqref{supmet} in the configuration space. As it can be
straightforwardly verified, the above given metric
$\bbar{G}_{\mu\nu}$ is flat and admits the following six Killing
vectors: \be \label{Killing}
\begin{aligned}
&\xi_1=\partial_f  ,\quad  \xi_2  = \frac{1}{2\, a\, b}\, \partial_a ,  \quad \xi_3=\frac{f}{2\,a\,b}\,\partial_a + \frac{1}{2\, b}\, \partial_f, \quad \xi_4   = -a\,\partial_a + b\,\partial_b - f \,\partial_f  ,\\
&\xi_5  = -\frac{a^2+ f^2}{2\,a\,b}\,\partial_a +\partial_b - \frac{f}{b}\,\partial_f  ,\quad \xi_6  = -a\, f\,\partial_a + b\, f \,\partial_b -\frac{a^2+f^2}{2}\,\partial_f.
\end{aligned}
\ee
These form an algebra under the Lie bracket; the non vanishing structure constants of this algebra are
\begin{align*}
& C^2_{31}=-C^2_{13}=C^1_{14}=-C^1_{41}=C^4_{61}=-C^4_{16}=1 \\
& C^2_{24}=-C^2_{42}=C^3_{26}=-C^3_{62}=C^5_{45}=-C^5_{54}=C^6_{46}=-C^6_{64}=1 \\
& C^3_{15}=-C^3_{51}=2,\quad C^5_{63}=-C^5_{36}=\frac{1}{2}.
\end{align*}
Additionally, the supermetric $\bbar{G}_{\mu\nu}$ exhibits a
homothetic symmetry
($\pound_{\xi_h}\bbar{G}_{\mu\nu}=\bbar{G}_{\mu\nu}$) generated by
\be \label{rnhom} \xi_h= \frac{a}{4} \frac{\partial}{\partial a} +
\frac{b}{4} \frac{\partial}{\partial b}+ \frac{f}{4}
\frac{\partial}{\partial f}, \ee which will be used in order to
completely integrate the system of the Euler-Lagrange equations
\begin{subequations} \label{eullag}
\begin{align}
(EL)_{N} & := -\frac{\partial L}{\partial N}, \\
(EL)_{q^i} & := \frac{d}{d r}\left(\frac{\partial L}{\partial \dot{q}^i }\right)-\frac{\partial L}{\partial q^i} \, .
\end{align}
\end{subequations}

Let us now turn to the Hamiltonian formulation; invoking the usual definition of the momenta
\begin{subequations} \label{momenta}
\begin{align} \label{momn}
p_N & := \frac{\partial L}{\partial N'}=0, \\ \label{moma} p_a & :=
\frac{\partial L}{\partial a'}= \frac{8\, a\, b\, b'}{N}, \\
\label{momb} p_b & := \frac{\partial L}{\partial b'} =\frac{8\, a \,
\left(b\, a'+a\, b'\right)}{N},\\ \label{momf} p_f & :=
\frac{\partial L}{\partial f'} = \frac{8\, b^2 \, f'}{N},
\end{align}
\end{subequations}
and following Dirac's algorithm \cite{Dirac}, we acquire one first
class primary constraint $p_N \approx 0$, the Hamiltonian \be
\label{hamiltonian} H = N\, \mathcal{H} = N \,
\left(-\frac{p_a^2}{16\, b^2}+\frac{p_a\, p_b}{8\, a\, b} +
\frac{p_f^2}{16\, b^2}-1\right), \ee and the first class secondary
constraint $\{p_N, H\}\approx 0\Rightarrow \mathcal{H}\approx 0$. If
we associate the phase-space quantities $Q_I:=\xi_I^\mu p_\mu$ with
the six Killing vector fields \eqref{Killing}, we are provided with
six linear integrals of motion \be \label{intofmo}
\begin{aligned}
&Q_1  = p_f  ,\quad  Q_2  = \frac{1}{2\, a\, b}\, p_a ,  \quad Q_3  = \frac{f}{2\,a\,b}\,p_a + \frac{1}{2\, b}\, p_f, \quad Q_4   = -a\,p_a + b\,p_b - f \,p_f  ,\\
&Q_5  = -\frac{a^2+ f^2}{2\,a\,b}\,p_a +p_b - \frac{f}{b}\,p_f  ,\quad Q_6  = -a\, f\,p_a + b\, f \,p_b -\frac{a^2+f^2}{2}\,p_f \, ,
\end{aligned}
\ee which form a Poisson bracket algebra with the previously
mentioned structure constants. As also stated in \cite{tchris1},
under the given lapse parametrization (in which the potential is
constant) the Poisson brackets of the $Q_I$'s with the Hamiltonian
$H$ are exactly equal to zero and not just weakly vanishing,
$\{Q_I,H\}=0$, for $I=1,\dots 6$. Moreover, since
$\mathcal{H}\approx 0$, the constancy of the potential part is
carried over to the quadratic in the momenta kinetic term, leading
inevitably the latter to become a Casimir invariant of the Lie
algebra formed by the $Q_I$'s. In the case we are studying this is
\be \label{casimir} Q_C = \frac{1}{4}\, \left(Q_2\, Q_5 +
Q_3^2\right)= \mathcal{H}+1. \ee

As it is known, the integrals of motion, $Q_I$'s, become constants,
say $\kappa_I$'s, on the solution space. However, these are not the
only existing integrals of motion. As shown in \cite{tchris2}, in
principle, all conformal Killing vectors of the supermetric define
rheonomic integrals of motion. For example, the relation $\pound_\xi
\bbar{G}_{\mu\nu}=\omega\, \bbar{G}_{\mu\nu}$ implies that if we
define the phase-space quantity $Q_\xi=\xi^\mu\, p_\mu$, then \be
\frac{dQ_\xi}{dr}=\{Q_\xi,H\} = \omega(q) \, \frac{N}{2}\,
\bbar{G}^{\mu\nu}\, p_\mu\, p_\nu = \omega(q) \, N, \ee holds. The
latter equality is valid since $\mathcal{H}=\frac{1}{2}\,
\bbar{G}^{\mu\nu}\, p_\mu\, p_\nu-1 \approx 0$. Thus, by integration
over $r$ the above equation is turned into the rheonomic integral
\be \label{genint} Q_\xi -\int\!\! \omega(q(r)) \, N dr=const. \ee
For $\omega=0$, there is no explicit $r$-dependence and the
corresponding integrals are just the $Q_I$'s generated by the six
Killing vector fields. In the case of a non vanishing $\omega$, the
usefulness of \eqref{genint} is limited, since one needs to know a
priori the trajectories $q(r)$ that solve the Euler-Lagrange
equations \eqref{eullag}. Nevertheless, for the homothetic Killing
field the previous problem is circumvented since $\omega=constant$;
another choice would be to pick up a particular conformal Killing
vector field and properly gauge fix the lapse, i.e. choose
$N=\frac{1}{\omega}$. In what follows we will use the homothetic
vector field $\xi_h$ and avoid any gauge fixing of the lapse $N$. We
thus write the following seven relations, that are valid on the
solution space:
\begin{subequations}
\begin{align} \label{autint}
Q_I& = \kappa_I, \quad \quad I=1,\ldots,6 \\ \label{rehint}
Q_h - \int\!\! N dr &= c_h \Rightarrow \frac{1}{4}(a\, p_a+ b\, p_b + f\, p_f) - \int\!\! N dr = c_h,
\end{align}
\end{subequations}
with $\kappa_I$'s and $c_h$ being constants. It is quite interesting
that the above relations completely determine the entire classical
solution space along with the two relations quadratic in the
$\kappa_I$'s emanating from the two Casimir invariants of the
algebra. Indeed, after substitution of \eqref{momenta}, if we choose
to algebraically solve the system of equations consisting of
\eqref{autint} for $I=1,\ldots,5$ and \eqref{rehint} with respect to
$a(r)$, $a'(r)$, $f(r)$, $f'(r)$, $\int\!\!N dr$ and $N(r)$, we
obtain the relations
\begin{subequations} \label{rnsol}
\begin{align}
a &= \pm \frac{\sqrt{-4\, b\, (\kappa_1\, \kappa_3+\kappa_2\, \kappa_4)+4\, b^2\, \left(\kappa_2\, \kappa_5+\kappa_3^2\right)+\kappa_1^2}}{2\, \kappa_2 \, b}, \\
a' &=  \mp \frac{b'\, \left(\kappa_1^2-2\, b\, (\kappa_1\, \kappa_3+ \kappa_2\, \kappa_4)\right)}{2 \, \kappa_2\, b^2 \sqrt{-4\, b \, (\kappa_1 \,\kappa_3+\kappa_2\, \kappa_4)+4\, b^2\, \left(\kappa_2 \, \kappa_5+\kappa_3^2\right)+\kappa_1^2}}, \\
f &= \frac{\kappa_3}{\kappa_2}-\frac{\kappa_1}{2\, \kappa_2 \, b}, \\
f' & = \frac{\kappa_1\, b'}{2\, \kappa_2\, b^2}, \\
\int\!\!N\, dr & = -\frac{-4\, b\, \left(\kappa_2\, \kappa_5+
\kappa_3^2\right)+4 \, c_h\, \kappa_2+2\, \kappa_1\, \kappa_3+3\,
\kappa_2\, \kappa_4}{4 \, \kappa_2}, \\ \label{rnsollap}
N&=\frac{4\, b'}{\kappa_2},
\end{align}
\end{subequations}
with $b$ remaining an arbitrary function of $r$. The consistency
conditions $a'=\frac{da}{dr}$ and $f'=\frac{df}{dr}$ are identically
satisfied, while $N=\frac{d}{dr}\int\!\!N dr$ leads to the
requirement \be \label{condcon} \kappa_2\, \kappa_5 + \kappa_3^2 =4,
\ee which is valid due to the Casimir invariant \eqref{casimir},
since the Hamiltonian (a.k.a. the quadratic constraint) is zero.
Additionally, and somewhat unexpectedly, if one substitutes
\eqref{rnsol} into the equation we have not used, i.e. $Q_6 =
\kappa_6$, one is  led to the following relation between constants:
\be \label{condcas2} \kappa_1\, \kappa_5 + 2\, \kappa_2\, \kappa_6 -
2\, \kappa_3\, \kappa_4 =0. \ee This relation is also valid on the
solution space, because of the existence of the second Casimir
invariant \be \label{secondcas} \widetilde{Q}_C = Q_1\, Q_5 + 2\,
Q_2\, Q_6 - 2\, Q_3\, Q_4. \ee If the form of the $Q_I$'s
\eqref{intofmo} is substituted into $\widetilde{Q}_C$ we find that
it vanishes identically, irrespectively of the classical solution.
Therefore, equation \eqref{condcas2} is retrieved on the solution
space .

It is an easy task to check that \eqref{rnsol} together with
\eqref{condcon} is the solution of the equations of motion
\eqref{eullag}. By a convenient reparametrization of the constants
$\kappa_I$ (four of which are arbitrary because of the requirements
\eqref{condcon} and \eqref{condcas2}) \be \label{kappas}
\begin{aligned}
\kappa_1&= -4\, Q, & \quad \kappa_2&= \frac{2}{c},& \quad \kappa_3&= \frac{2\, c_3}{c}, \\
\kappa_4&= 4\, c\, m+c_3\, Q,& \quad \kappa_5&=2\, c-\frac{2\,
c_3^2}{c},& \quad \kappa_6&= 2\, \left(Q \,
\left(c^2+c_3^2\right)+2\, c\, c_3\, m\right),
\end{aligned}
\ee the corresponding space-time line element in \eqref{metric}
turns out to be \be \label{rnmetb} ds^2= - c^2\,\left(1-\frac{2\,
m}{b(r)}+\frac{Q^2}{b^2(r)}\right)dt^2 + \left(1-\frac{2\, m}{b(r)}+
\frac{Q^2}{b^2(r)}\right)^{-1}db^2(r)+ b^2(r)\, d\theta^2 + b^2(r)
\, \sin^2\theta\, d\phi^2 \ee which, of course, is the well known RN
metric  involving only two essential parameters: the mass $m$ and
the charge $Q$; $c$ is absorbable by a re-scaling of the time
coordinate, i.e. $t\rightarrow \frac{t}{c}$ \be \label{rnmet}
ds_{RN}^2= - \left(1-\frac{2\, m}{r}+\frac{Q^2}{r^2}\right)dt^2 +
\left(1-\frac{2\, m}{r}+ \frac{Q^2}{r^2}\right)^{-1}dr^2+ r^2\,
d\theta^2 + r^2 \, \sin^2\theta\, d\phi^2 . \ee Some remarks are in
order:
\begin{itemize}
\item $\kappa_1$, and therefore $Q_1$, is the only linear integral of motion depending solely on an essential constant.
\item The quantities $Q_2$, $Q_3$ and $Q_5$ incorporate the non essential constants $c$ and $c_3$.
Therefore, these can be claimed to be completely gauged fixable,
since one can utilize the arbitrariness of $c$ and $c_3$ to change
their values.
\item $Q_4$ and $Q_6$ depend on both essential and non essential constants, but still their values are gauge dependent.
\item The value of the Casimir invariant, $Q_C$, on the solution space is:
\be \label{cascon} \frac{1}{4}\,(\kappa_2\,\kappa_5+\kappa_3^2) =1,
\ee as expected by \eqref{casimir}, since the Hamiltonian constraint
$\mathcal{H}$ is weakly zero.
\end{itemize}
We can add here, that the constant $c$ can be set equal to one but
not zero. On the other hand, $c_3$ can be taken equal to zero since
it is absorbed additively. Moreover, $c_3$ is connected to the gauge
freedom of the electrostatic scalar potential, since by using
\eqref{kappas} one can see that $f(r)=c_3-\frac{c\, Q}{b(r)}$.

By setting $c=1$, and $c_3=0$ the values of the six $\kappa_I$'s
become \be \label{kapfix} \kappa_1=-4\,Q , \quad \kappa_2=2, \quad
\kappa_3=0, \quad \kappa_4=4\, m, \quad \kappa_5= 2, \quad
\kappa_6=2\, Q \ee which are the values one would obtain if solution
\eqref{rnmet} had been taken as the starting point for the
computation of the linear integrals of motion.

\section{Quantization through symmetries} \label{sect3}
The identification of the linear integrals of motion as physical
quantities leads to the need of expressing them as operators. The
algebra defined by these operators has to match the classical Lie
algebra and, moreover, one has to determine which of them can be
applied at the same time on the wave function together with the
constraints mentioned in the previous section. These issues have
been clearly addressed in \cite{tchris1}. In a quick view, we start
with the usual definition of the momenta ($\hbar=1$) as operators
\be p_\alpha \rightarrow \widehat{p}_\alpha := -\ima
\frac{\partial}{\partial q^{\alpha}}, \ee where $q^{\alpha}$ is any
one of $a$, $b$, $f$, $N$. After that, the quantum analogues of the
conditional symmetries $Q_I$, are expressed in the most general form
of a linear Hermitian (under an arbitrary measure $\mu$)
differential operator of the first order: \be \label{qlin}
\widehat{Q}_I := - \frac{\ima}{2\mu} \left(\mu \, \xi_I^\alpha\,
\partial_\alpha + \partial_\alpha \, \mu\, \xi_I^\alpha \right). \ee
It has been proved in \cite{tchris1} that operators $\widehat{Q}_I$
defined as in \eqref{qlin} satisfy the same algebra as do the
classical quantities $Q_I$, i.e.
$[\widehat{Q}_I,\widehat{Q}_J]F=C^K_{IJ}\,\widehat{Q}_KF$ for any
function $F$ for which the action of the operators is well defined.
It is noteworthy that this happens for any arbitrary measure
$\mu(a,b,f)$.

Apart from the primary constraint \be \widehat{p}_N=-\ima
\frac{\partial}{\partial N} \Psi=0 \Rightarrow \Psi=\Psi(a,b,f), \ee
the main operator  one has to apply is the quantum analogue of the
Hamiltonian constraint or, equivalently in the particular lapse
parametrization, of the Casimir invariant ($\widehat{Q}_C$), since
\be \label{hamcasrel} \widehat{\mathcal{H}}\Psi = (\widehat{Q}_C
-1)\Psi=0. \ee In order to fix the kinetic part of the Hamiltonian
operator we demand Hermiticity under the same measure $\mu$; we thus
have \cite{tchris3} \be \label{hamoperator} \widehat{\mathcal{H}}_c
\Psi = \left[- \frac{1}{2\mu} \partial_\alpha (\mu \,
\bbar{G}^{\alpha\beta} \partial_\beta) - 1 \right] \Psi = 0. \ee The
addition of a term proportional to the Ricci scalar of the
supermetric $\bbar{G}_{\alpha\beta}$ is not needed since the
superspace is flat. In what follows we will, invoking a sense of
naturality, choose the measure $\mu$ to be equal to $\sqrt{|\det
\bbar{G}_{\alpha\beta}|}=16\sqrt{2}\, a\, b^2$. This choice ensures
that the derivative part of the quadratic constraint operator
becomes the Laplace-Beltrami operator which is also scalar under
general configuration space transformations. Further, it also
renders the linear operators \eqref{qlin} pure derivations, i.e. it
makes them have the derivatives acting on the far right since
$(\mu\,\xi^\beta_I)_{;\beta}$ for every $I=1...6$.

Apart from $\widehat{\mathcal{H}}$, we also have at our disposal the
conditional symmetries. They can too act on the wave function and
provide the connection to the solution space of the classical
theory. The wave function of the system is to be realized as an
eigenstate of those physical quantities that can be measured
together:
 \be \widehat{Q}_I\Psi=\kappa_I\Psi , \ee
 for all the subsets of $Q_I$'s for which the structure constants of the
subalgebra they form, satisfy the integrability conditions \be
\label{intcon} C^I_{JK}\kappa_I=0. \ee Equation \eqref{intcon} has
been proven as an integrability condition in \cite{tchris1},
\cite{tchris4} and gives a selection rule for determining those
operators which can be applied at the same time on the wave
function. The results of the use of \eqref{intcon} can be
summarized, according to the various subalgebras, as follows:
\begin{enumerate}
\item For the entire algebra and for all five and four dimensional subalgebras \eqref{intcon} is not valid.
\item For the non Abelian three dimensional subalgebra $\{Q_1,Q_4,Q_6\}$, the integrability condition \eqref{intcon}
implies that all the corresponding $\kappa_I$, $I=1,4,6$ must be
zero (since the algebra is semi-simple). For a generic configuration
this is unacceptable in view of the fact that, for instance,
$\kappa_1$ corresponds to the essential constant $Q$.
\item For the three non Abelian two dimensional subalgebras $\{Q_2,Q_4\}$, $\{Q_4,Q_5\}$ and $\{Q_4,Q_6\}$, the results of the application of \eqref{intcon} are similar to the previous case. For the first of them \eqref{intcon} implies that $\kappa_2=0$, a condition that cannot be met in view of $\kappa_2=\frac{2}{c}$ (see \eqref{kappas}). For the other two $\kappa_5$ or $\kappa_6$ respectively must be zero, a fact implying a kind of gauge fixing for the constants $c$ and $c_3$, hence restricting the generality.
\end{enumerate}
We are thus led to consider the following Abelian subalgebras:
\begin{enumerate}
\item the three dimensional subalgebra made up by $Q_2$, $Q_3$ and $Q_5$
\item the two dimensional subalgebras:
\begin{enumerate}
\item $Q_1$, $Q_2$ \label{2a}
\item $Q_2$, $Q_3$ \label{2b}
\item $Q_2$, $Q_5$ \label{2c}
\item $Q_3$, $Q_4$  \label{2d}
\item $Q_3$, $Q_5$ \label{2e}
\item $Q_5$, $Q_6$  \label{2f}
\end{enumerate}
\end{enumerate}
Of course, there are also six one dimensional subalgebras but these
cannot be considered on account of the existence of \emph{two}
essential constants needed to describe the underlying geometry.
Cases \eqref{2a}, \eqref{2d} and \eqref{2f} of the two dimensional
subalgebras are of particular interest, since they involve integrals
that are connected with essential constants (those are $Q_1$, $Q_4$
and $Q_6$). Let us proceed with the examination of each case.

\subsection{The three dimensional subalgebra and the marginal cases $\eqref{2b}$, $\eqref{2c}$ and $\eqref{2e}$}
In considering the three dimensional Abelian subalgebra spanned by $Q_2$, $Q_3$ and $Q_5$, and with the choice of
measure $\mu=16\sqrt{2}\, a\, b^2$, the given $\xi_I$'s in \eqref{Killing} and definitions \eqref{qlin}, we obtain
the following set of differential equations:
\begin{subequations}
\begin{align} \label{Q2}
&\widehat{Q}_2\Psi = \kappa_2\, \Psi \Rightarrow \frac{\ima}{2\, a\,
b} \partial_a\Psi+\kappa_2\Psi=0, \\ \label{Q3} &\widehat{Q}_3\Psi =
\kappa_3\, \Psi \Rightarrow \ima \left(\frac{f}{2\, a\,
b}\partial_a\Psi+\frac{1}{2\, b}\partial_f\Psi\right)+\kappa_3\,
\Psi =0,\\ \label{Q5} &\widehat{Q}_5\Psi = \kappa_5\, \Psi
\Rightarrow \ima \left[\left(\frac{a^2+f^2}{2\, a\,
b}\right)\partial_a\Psi-\partial_b\Psi+\frac{f}{b}\partial_f\Psi\right]-\kappa_5\,
\Psi=0,
\end{align}
\end{subequations}
together with the Hamiltonian constraint
\be \label{WDW}
\widehat{\mathcal{H}}\Psi = \frac{1}{8\,
b}\left[\frac{1}{2\,b}\left(\partial_{aa}\Psi-\partial_{ff}\Psi\right)-\frac{1}{a}\partial_{ab}\Psi\right]-\Psi=0.
\ee
By solving successively from \eqref{Q2} to \eqref{Q5}, the
dependence of $\Psi(a,b,f)$ on its arguments is completely
determined:
\be \label{sol1} \Psi= \lambda\, e^{\ima\,
b\left(\kappa_2\left(a^2-f^2\right)+2\, \kappa_3\, f +
\kappa_5\right)},
\ee
with $\lambda$ being an arbitrary constant. By
substituting solution \eqref{sol1} into \eqref{WDW} we get \be
\kappa_2\, \kappa_5 +\kappa_3^2 - 4 =0, \ee which is an identity in
view of \eqref{cascon}.

The state of the system described by \eqref{sol1} resembles the
situation that arose in \cite{tchris1} for the case of Schwarzschild
geometry. There too, the enforcement of the maximal Abelian subgroup
led to a plane wave solution. Furthermore, that algebra was also
spanned by integrals of motion which had no connection to essential
constants of the underlying geometry.

If we now choose to consider the two dimensional cases that are made up from $Q_2$, $Q_3$ and $Q_5$,
namely $\eqref{2b}$, $\eqref{2c}$ and $\eqref{2e}$,
we are led to essentially the same solution for $\Psi$.
\begin{itemize}
\item The set of equations \eqref{Q2}, \eqref{Q3} and \eqref{WDW} leads to a solution that differs from \eqref{sol1} by a phase $\frac{\kappa_2\, \kappa_5 +\kappa_3^2 - 4}{\kappa_2}$ which, however, is zero due to \eqref{cascon}.
\item If we now consider equations \eqref{Q2}, \eqref{Q5} and \eqref{WDW}, we end up with the following wave
function:
\be \label{sol25} \Psi_{25}= \lambda_1
\,e^{2\,b\,f\sqrt{\kappa_2\,\kappa_5-4}}\, e^{\ima\,
b\left(\kappa_2\left(a^2-f^2\right)+ \kappa_5\right)}+\lambda_2
\,e^{-2\,b\,f\sqrt{\kappa_2\,\kappa_5-4}}\,e^{\ima\,
b\left(\kappa_2\left(a^2-f^2\right)+ \kappa_5\right)},
\ee
that seems quite different from \eqref{sol1}. Nevertheless, as we have
previously mentioned, there is a non essential constant (the constant $c_3$ which refers to the freedom of the scalar potential $f(r)$)
that can be set to zero by a gauge transformation. Then,
$\kappa_3$ becomes zero and \eqref{cascon} leads to
$\kappa_2\,\kappa_5=4$. Under this condition, \eqref{sol25} becomes
\eqref{sol1}.
\item Lastly, we take into account the set of equations \eqref{Q3}, \eqref{Q5} and \eqref{WDW}.
The common solution of this set is different from \eqref{sol1} by a phase
$\frac{a^2-f^2}{\kappa_5}\left(\kappa_2\, \kappa_5 +\kappa_3^2 - 4\right)$,
which again is zero because of \eqref{cascon}.
\end{itemize}
So, as it is evident from the above considerations, all three cases are connected to each other, giving the same
plane wave solution that emerges from the consideration of the maximal Abelian algebra.

\subsection{The two dimensional case \eqref{2a} ($Q_1$, $Q_2$)}
This subalgebra contains $Q_1$ whose value on the solution space is
proportional to the essential constant $Q$, ($\kappa_1=-4\,Q$),
meaning that $Q_1$ is purely connected to a quantity referring to
the geometry of space-time. We consider equation \be \label{Q1}
\widehat{Q}_1\Psi = \kappa_1\, \Psi \Rightarrow \ima\,\partial_f
\Psi+\kappa_1\,\Psi=0, \ee together with \eqref{Q2} and \eqref{WDW}.
The common solution for the given set of equations is \be
\label{sol2} \Psi =\frac{\lambda}{\sqrt{b}}\,
\exp\left(\ima\frac{\kappa_1^2+4\, b\, f \kappa_1\,\kappa_2+4\, a^2
b^2\kappa_2^2+16\, b^2}{4\, b\, \kappa_2}\right), \ee with $\lambda$
being again an arbitrary constant. With this wave function we are
led to a probability density \be \mu\, \Psi^*\, \Psi \propto a\, b,
\ee that encompasses only the two scale factors and is completely
free of the variable $f$. The latter is only present in the phase of
the wave function.

\subsection{The two dimensional case \eqref{2d} ($Q_3$, $Q_4$)}
The linear integral $Q_4$ assumes the constant value $\kappa_4=4\,
(c\, m+c_3\, Q)$ on the solution space. As we can see, it bears a
connection mainly to $m$, since $c_3$ can be set equal to zero. However,
its value, in contrast to the previous case, is somewhat gauge
dependent due to the involvement of non essential constants. In this
case we use equation
\be \label{Q4} \widehat{Q}_4\Psi = \kappa_4\,
\Psi \Rightarrow \ima\,\left(a\,\partial_a \Psi -b\,\partial_b\Psi+
f\,\partial_f\Psi\right)-\kappa_4\,\Psi=0,
\ee
as well as \eqref{Q3} and the WDW equation \eqref{WDW}. The integration of \eqref{Q4}
leads to a solution of the form
\be
\Psi(a,b,f)=a^{-\ima\,\kappa_4}\psi_1(a\, b , \frac{f}{a}).
\ee
It is useful to use the new variables $u=b\, a$, $v=\frac{f}{a}$ and
$a$, for which the imposition of equation \eqref{Q3} on the previous
wave function leads to \be \label{Q3uv}
\ima\,\left((v^2-1)\,\partial_v\psi_1+u\,
v\,\partial_u\psi_1\right)+(2\kappa_3\, u +\kappa_4\, v)\psi_1. \ee
Even though $\kappa_3$ can be set equal to zero through a gauge
transformation, we choose to carry it until the final result. The
solution of \eqref{Q3uv} reads \be \psi_1(u,v)
=e^{2\,\ima\,\kappa_3\,u\,v}\, u^{\ima \, \kappa_4}
\psi_2(\ln(u\sqrt{v^2-1})). \ee At this stage, a new change of
variables is in order; setting $u=\frac{e^w}{\sqrt{v^2-1}}$ the WDW
equation \eqref{WDW} becomes \be
\psi_2''(w)+2\,\ima\,\kappa_4\,\psi_2'(w)+4\,e^{2\, w}\,
(\kappa_3^2-4)\psi_2(w)=0. \ee The general solution of this equation
is \be
\begin{aligned}
\psi_2(w)=e^{-\frac{1}{2} \,\kappa_4 (\pi +2 \,\ima\, w)} \left[ \lambda_1\, I_{\ima\, \kappa_4}\left(2\,e^{w} \sqrt{\left(\kappa_3^2-4\right)}\right)+ \right. \\
\left.  \lambda_2 \, I_{-\ima \,\kappa_4}\left(2\,e^{w}
\sqrt{\left(\kappa_3^2-4\right)}\right)\right],
\end{aligned}
\ee with $\lambda_1$, $\lambda_2$ being arbitrary constants while $I_\nu(x)$ is
the modified Bessel function of the first kind. Thus, the final form
of the wave function $\Psi(a,b,f)$ is: \be\label{sol3}
\begin{aligned}
\Psi=\left(a^2-f^2\right)^{-\frac{1}{2} (\ima \, \kappa_4)} e^{2\,\ima\, b\, f \,\kappa_3} \left[\lambda_1 \, I_{\ima \,\kappa_4}\left(2\, b\, \sqrt{a^2-f^2} \sqrt{\left(\kappa_3^2-4\right)}\right)+\right. \\
\left.\lambda_2\, I_{-\ima \,\kappa_4}\left(2\, b\, \sqrt{a^2-f^2}
\sqrt{\left(\kappa_3^2-4\right)}\right)\right].
\end{aligned}
\ee

\subsection{The two dimensional case \eqref{2f} ($Q_5$, $Q_6$)}
The constant value of $Q_6$ is $\kappa_6=2\,\left(2\, c_3\, c \,
m+(c_3^2 \, +c^2 )\, Q\right)$. Under the gauge conditions $c_3=0$
and $c=1$, $\kappa_6$ equals to $2\, Q$. Our starting point is the
differential equation \be \label{Q6} \widehat{Q}_6\Psi = \kappa_6\,
\Psi \Rightarrow
\ima\,\left(a\,f\,\partial_a\Psi-b\,f\,\partial_b\Psi+\frac{1}{2}(a^2+f^2)\,\partial_f\Psi\right)-\kappa_6\,\Psi=0,
\ee whose solution is \be \Psi(a,b,f)=e^\frac{-2\,\ima\,
\kappa_6\,f}{a^2-f^2}\, \psi_1 (a\, b, \frac{f^2}{a}-a). \ee By
defining as new variables $u=a\, b$ and $v=\frac{f^2}{a}-a$ and
substituting the above form of $\Psi$ in equation \eqref{Q5} we get
\be \ima\, v \,
\left(v\,\partial_v\psi_1(u,v)-u\,\partial_u\psi_1(u,v)\right)+2\,\kappa_5\,
u\, \psi_1(u,v)=0. \ee Its integration yields the
function \be \psi_1(u,v)=e^{-\ima\frac{\kappa_5\, u}{v}}\psi_2(u\,
v). \ee At this stage we introduce the new variable $w=u\, v$.
Subsequent substitution into the WDW equation \eqref{WDW} leads to
\be 2\, \ima\,\kappa_5\, w^2\,
\psi_2'(w)+(2\,\kappa_6^2+(\ima\,\kappa_5-8\,w)\,w)\psi_2(w)=0, \ee
admitting the solution \be \psi_2(w)
=\frac{\lambda}{\sqrt{w}}\exp\left(-\ima\frac{\kappa_6^2+4\,w^2}{\kappa_5\,w}\right).
\ee The wave function is written in the original variables as \be
\label{sol4} \Psi(a,b,f) =
\frac{\lambda}{\sqrt{b\,(f^2-a^2)}}\exp\left(\frac{\ima\, \left(4\,
a^4\, b^2 + 4\, b^2\, f^4 + a^2\, b^2\, (-8\, f^2 + \kappa_5^2) - 2
\,b\, f\, \kappa_5\, \kappa_6 +
   \kappa_6^2\right)}{b\, (a^2 - f^2) \,\kappa_5}\right),
\ee
and leads to a probability density
\be
\mu\,\Psi^*\,\Psi \propto \frac{a\, b}{f^2-a^2}.
\ee

At this point, one could think that we have attained two different
representations for the physical quantity $Q$: The first was the
case \eqref{2a} with the use of $\widehat{Q}_1$ and $\widehat{Q}_2$,
where classically $\kappa_1=4\, Q$. The second is this, with
$\widehat{Q}_5$ and $\widehat{Q}_6$ (under gauge conditions
$c=1$, $c_3=0$, $\kappa_6=2\, Q$).

However, the wave function
\eqref{sol4}, under the transformation
$(a,b,f)\rightarrow(\alpha,\beta,\phi)$  with \be
a=\frac{\alpha}{\alpha^2-\phi^2}\quad, \quad
b=\beta\,(\phi^2-\alpha^2)\quad, \quad
f=\frac{\phi}{\phi^2-\alpha^2}, \ee and $\kappa_5$, $\kappa_6$
expressed in the gauge $c=1$, $c_3=0$, is turned into \be
\Psi(\alpha,\beta,\phi)=\frac{\lambda}{\sqrt{\beta}}\exp\left(\frac{-2\ima\,\left((1+\alpha^2)\,\beta^2
-2\,Q\,\beta\,\phi+Q^2\right)}{\beta}\right). \ee In the same gauge,
the wave function \eqref{sol2} becomes \be
\Psi_{12}(a,b,f)=\frac{\lambda}{\sqrt{b}}\exp\left(\frac{-2\ima\left((1+a^2)\,b^2
-2\,Q\,b\,f+Q^2\right)}{b}\right). \ee These two wave functions
assume the same functional form. What is important though is, that the very same transformation transforms the Killing vector of the
supermetric $\xi_6$ into $\frac{1}{2}\xi_1$ in the new variables
(the factor $\frac{1}{2}$ expresses the fact that $\kappa_1=4\,
Q$ while $\kappa_6=2\, Q$ under the considered gauge).


\section{Semiclassical analysis} \label{sect4}
In this section we are going to present a semiclassical analysis of
the problem reviewed in the previous sections. To accomplish this
task, we examine a wave function of the form

\begin{equation}\label{A}
\Psi(a,b,f)=\Omega(a,b,f)e^{\ima S(a,b,f)},
\end{equation}in the WDW equation \eqref{WDW}. Here $\Omega(a,b,f)$ and $S(a,b,f)$ are
some real functions representing the magnitude and the phase of the
wave function, respectively. Upon using this expression for the wave
function, the WDW equation leads to the continuity equation
\begin{equation}\label{B}
\frac{1}{16b^2}\left[2\left(\frac{\partial \Omega}
{\partial a}\frac{\partial S}{\partial a}-\frac{\partial \Omega}{\partial f}\frac{\partial S}
{\partial f}\right)+\Omega \left(\frac{\partial^2 S}{\partial a^2}-\frac{\partial^2 S}{\partial f^2}\right)\right]
-\frac{1}{8ab}\left(\frac{\partial \Omega}{\partial a}\frac{\partial S}{\partial b}+\frac{\partial \Omega}{\partial b}
\frac{\partial S}{\partial a}+\frac{\partial^2 S}{\partial a \partial b}\right)=0,
\end{equation}
and the modified Hamilton-Jacobi equation
\begin{equation}\label{C}
-\frac{1}{16b^2}\left(\frac{\partial S}{\partial a}\right)^2+\frac{1}{8ab}\frac{\partial S}{\partial a}\frac{\partial
S}{\partial b}+\frac{1}{16b^2}\left(\frac{\partial S}{\partial f}\right)^2-1+{\cal Q}=0,
\end{equation}
in which
\begin{equation}\label{D}
{\cal Q}=\frac{1}{\Omega}\left[\frac{1}{16b^2}\left(\frac{\partial^2 \Omega}{\partial a^2}-\frac{\partial^2 \Omega}{\partial f^2}\right)-\frac{1}{8ab}\frac{\partial^2 \Omega}{\partial a\partial b}\right],
\end{equation}
is the quantum potential. A glance at equation (\ref{C}) shows that it is of the form
\begin{equation}\label{E}
{\cal H}\left(q^{\mu},p_{\mu}=\frac{\partial S}{\partial
q^{\mu}}\right)+{\cal Q}=0,
\end{equation}
where ${\cal H}$ is the Hamiltonian defined in \eqref{hamiltonian},
$q^{\mu}=(a,b,f)$ are the variables of the configuration space and
$p_{\mu}=(p_a,p_b,p_f)$ are the momenta conjugate to $q^{\mu}$ given
by \eqref{moma}-\eqref{momf}. Therefore, in the semiclassical
picture, the equations of motion can be written as

\begin{eqnarray}\label{G}
\left\{
\begin{array}{ll}
\dfrac{8}{N}abb'=\dfrac{\partial S}{\partial a},\\\\
\dfrac{8}{N}\left(aba'+a^2b'\right)=\dfrac{\partial S}{\partial b},\\\\
\dfrac{8}{N}b^2f'=\dfrac{\partial S}{\partial f}.
\end{array}
\right.
\end{eqnarray}
If the quantum potential (\ref{D}) is non-zero, the solutions to the
above system differ from the classical solutions by some correction
terms coming from the quantum mechanical considerations; in the
cases where the quantum potential is equal to zero, we expect that
solving the system (\ref{G}) will reproduce the pure classical
solutions. In the following subsections we will deal with this issue
with the help of the wave functions obtained in the previous
section.

\subsection{The three dimensional subalgebra and the two dimensional marginal cases}
We start with the wave function \eqref{sol1} which, with the notation introduced in this section, yields
\begin{equation}\label{H}
S(a,b,f)=b\left[\kappa_2(a^2-f^2)+2\kappa_3
f+\kappa_5\right],\hspace{5mm}\Omega=\mbox{const.}
\end{equation}
It is clear that the quantum potential is zero, hence nothing but the classical solutions may be retrieved by the
semiclassical analysis. Indeed, in this case the system (\ref{G}) takes the form
\begin{eqnarray}\label{I}
\left\{
\begin{array}{ll}
\dfrac{8}{N}abb'=2\kappa_2 ab,\\\\
\dfrac{8}{N}\left(aba'+a^2b'\right)=\kappa_2(a^2-f^2)+2\kappa_3f+\kappa_5,\\\\
\dfrac{8}{N}b^2f'=2b(\kappa_3-\kappa_2 f).
\end{array}
\right.
\end{eqnarray}
To solve the above system of equations, let us for the moment assume $N=2$
(this assumption will be justified later) while we use the numerical
values \eqref{kapfix} for the $\kappa_I$'s. Under these conditions, the
first equation of (\ref{I}) can be immediately integrated giving
\begin{equation}\label{J}
b(r)=r,
\end{equation}
in which we have ignored an additive integration constant. Using this result in the third equation of (\ref{I}) we
obtain
\begin{equation}\label{K}
f(r)=\frac{C_1}{r},
\end{equation}
where $C_1$ is an integration constant. Now, upon insertion of these expressions for $b(r)$ and $f(r)$ in the second equation of (\ref{I}) we arrive at the following differential equation for
$a(r)$:
\begin{equation}\label{L}
2ra(r)a'(r)+a^2(r)=1-\frac{C_1^2}{r^2},
\end{equation}
which admits the solution
\begin{equation}\label{M}
a(r)=\left(1+\frac{C_2}{r}+\frac{C_1^2}{r^2}\right)^{1/2},
\end{equation}
where $C_2$ is another constant of integration. A simple calculation based on the above relations gives
\begin{equation}\label{N}
2aba'b'+a^2b'^2+b^2f'^2=1,
\end{equation}
which shows that the assumption $N=2$ is compatible with the
expression \eqref{rnsollap} for the lapse function. Now, if we
identify the integration constants with the charge and mass
parameters as $C_1=Q$ and $C_2=-2m$, the line element \eqref{metric}
takes the form of a RN black hole \eqref{rnmet}, as expected in the
case of vanishing quantum potential.

\subsection{The two dimensional subalgebra $(Q_1,Q_2)$}
In this subalgebra the wave function is given by \eqref{sol2} for
which again we have used the numerical values
\eqref{kapfix} for the $\kappa_I$'s
\begin{equation}\label{P}
S(a,b,f)=\frac{2Q^2-4Qbf+2a^2b^2+2b^2}{b},\hspace{5mm}\Omega(a,b,f)=\frac{\lambda}{\sqrt{b}}.
\end{equation}
From (\ref{D}) it is seen that the quantum potential is
again equal to zero. The equations of the system (\ref{G}) become
\begin{eqnarray}\label{R}
\left\{
\begin{array}{ll}
\dfrac{8}{N}abb'=4ab,\\\\
\dfrac{8}{N}\left(aba'+a^2b'\right)=2+2a^2-\dfrac{2Q^2}{b^2},\\\\
\dfrac{8}{N}b^2f'=-4Q,
\end{array}
\right.
\end{eqnarray}
which, again after choosing $N=2$, can be easily integrated
providing the result
\begin{equation}\label{S}
b(r)=r,\hspace{5mm}f(r)=\frac{Q}{r}+C_1,\hspace{5mm}a(r)=\left(1+\frac{C_2}{r}+\frac{Q^2}{r^2}\right)^{1/2}.
\end{equation}
We see that the standard form \eqref{rnmet} of the classical RN
black hole solution  can be recovered if one sets the integration
constant $C_1=0$ and identifies the integration constant $C_2$ with the mass parameter as
$C_2=-2m$. It seems appropriate to mention that the solutions
(\ref{J}), (\ref{K}) and (\ref{M}) of the three dimensional
subalgebra do not contain any of the particular values of the
essential parameters of the RN black hole, but $Q$ and $m$ appear as
integration constants after solving the system. However, in the
solutions (\ref{S}) the charge parameter enters directly into the
space-time geometry (not as an integration constant) while the mass
parameter is still an integration constant. This is a reflection of
the fact that none of the constant values
$(\kappa_2,\kappa_3,\kappa_5)$ of the quantities $(Q_2,Q_3,Q_5)$
which span the three dimensional subalgebra depends on the essential
constants, while in the two dimensional case $(Q_1,Q_2)$, the
constant $\kappa_1$ is indeed essential.

\subsection{The two dimensional subalgebra $(Q_3,Q_4)$}
In this case, the expression \eqref{sol3} gives the wave function in
terms of the Bessel functions. However, since the Bessel functions
can be written as a superposition of the Hankel functions, we write
the wave function as
\begin{equation}\label{S1}
\Psi(a,b,f)=(a^2-f^2)^{-2\ima m}\left[c_1H_{4\ima m}^{(1)}\left(4b\sqrt{a^2-f^2}\right)+c_2H_{4\ima m}^{(2)}\left(4b\sqrt{a^2-f^2}\right)\right],
\end{equation}
where $H_{\nu}^{(1),(2)}(z)$ are the Hankel functions of the first
and second kind, respectively, and we have used the numerical values \eqref{kapfix}
for $\kappa_I$'s. In the classical limit, i.e. for large values of $r$,
we have $b(r)\sim r$, $a(r)\sim 1$ and $f(r)\sim 0$. Under these
conditions the argument of the aforesaid Hankel functions takes a large
value and therefore, in view of the asymptotical behavior of the
Hankel functions which is $H_{\nu}^{(1),(2)}(z)\sim z^{-1/2}e^{\pm
\ima [z-(2\nu+1)\pi/4]}$, we can infer the following form of the
wave function in the semiclassical approximation:
\begin{equation}\label{S2}
\Psi(a,b,f)\sim \frac{1}{\sqrt{b}(a^2-f^2)^{1/4}}
(a^2-f^2)^{-2\ima m}e^{4\ima b\sqrt{a^2-f^2}}.
\end{equation}
Hence, comparing this expression with (\ref{A}) we get
\begin{equation}\label{S3}
\Omega(a,b,f)\sim \frac{1}{\sqrt{b}(a^2-f^2)^{1/4}},
\end{equation}
and
\begin{equation}\label{S4}
S(a,b,f)=-2m \ln (a^2-f^2)+4b\sqrt{a^2-f^2}.
\end{equation}
From (\ref{S3}) and with the help of (\ref{D}) one obtains the
quantum potential
\begin{equation}\label{S5}
{\cal Q}(a,b,f)=-\frac{1}{64b^2(a^2-f^2)},
\end{equation}
thereby observing that, unlike the previous subsection, its value
is not equal to zero. Therefore, due to quantum effects, some
modifications  are expected to appear upon solving the
system of equations ({\ref{G}). Using the expression (\ref{S4}) this
system takes the form
\begin{eqnarray}\label{S6}
\left\{
\begin{array}{ll}
\dfrac{8}{N}abb'=\dfrac{4ab}{\sqrt{a^2-f^2}}-\dfrac{4ma}{a^2-f^2},\\\\
\dfrac{8}{N}\left(aba'+a^2b'\right)=4\sqrt{a^2-f^2},\\\\
\dfrac{8}{N}b^2f'=-\dfrac{4bf}{\sqrt{a^2-f^2}}+\dfrac{4mf}{a^2-f^2}.
\end{array}
\right.
\end{eqnarray}
If, as before, we choose the gauge $N=2$, the first and the third equations of the above system give $f'/f=-b'/b$
which can be immediately integrated to obtain
\begin{equation}\label{S7}
f(r)=\frac{Q}{b(r)},
\end{equation}
where $Q$ is an integration constant. With this relation at hand, after some algebra with the first and the second
equations of (\ref{S6}), we get
\begin{eqnarray}\label{S8}
\left\{
\begin{array}{ll}
\dfrac{a'}{a}=-\dfrac{Q^2}{a^2 b^2 \sqrt{a^2 b^2-Q^2}}+\dfrac{m}{a^2 b^2-Q^2},\\\\
\dfrac{b'}{b}=\dfrac{1}{\sqrt{a^2 b^2-Q^2}}-\dfrac{m}{a^2 b^2-Q^2},
\end{array}
\right.
\end{eqnarray}
which gives rise to
\begin{equation}\label{S9}
(ab)'=\frac{\sqrt{a^2b^2-Q^2}}{ab},
\end{equation}
from which we obtain
\begin{equation}\label{S10}
a^2b^2=r^2+Q^2,
\end{equation}
where a constant of integration has been set equal to zero. Now, with a straightforward calculation based on the system (\ref{S8})
and (\ref{S7}) we find
\begin{equation}\label{S11}
a(r)=e^{-m/r}\left(1+\frac{Q^2}{r^2}\right)^{1/2},\hspace{5mm}b(r)=re^{m/r},\hspace{5mm}f(r)=\frac{Q}{r}e^{-m/r}.
\end{equation}
Again we see that the essential constant $m$ enters, in this case,
directly into the space-time metric while the essential matter
parameter $Q$ appears as an integration constant. The solutions
(\ref{S11}) tend asymptotically to the RN line-element
\eqref{rnmet}, however, unlike the RN solution, this one does not exhibit a
horizon-like singularity. Now, let us see what happens in the limit
of small $r$. In this limit the argument of the Bessel functions in
the wave function \eqref{sol3} is small. According to the behavior
$z^{\nu}(\lambda_1+\lambda_2 z^2+O(z^4))$ for the Bessel function
with a small argument, the wave function takes the form
\begin{equation}\label{S12}
\Psi(a,b,f)=\left[\lambda_1+\lambda_2b^2(a^2-f^2)\right]e^{4\ima m\ln
b},
\end{equation}
which, with the notation of (\ref{A}), gives
\begin{equation}\label{S13}
\Omega(a,b,f)=\left[\lambda_1+\lambda_2b^2(a^2-f^2)\right],
\end{equation}
and
\begin{equation}\label{S14}
S(a,b,f)=4m\ln b.
\end{equation}
Expression (\ref{S13}) yields a non-zero quantum potential of the form
\begin{equation}\label{S15}
{\cal
Q}(a,b,f)=-\frac{\lambda_2}{4\left[\lambda_1+\lambda_2b^2(a^2-f^2)\right]}
\end{equation}
while with (\ref{S14}) the system {(\ref G}) admits the solution
\begin{equation}\label{S16}
a(r)=(2mr+a_0)^{1/2},\hspace{5mm}b(r)=\beta\,(\mbox{const.}),\hspace{5mm}f(r)=\mbox{const.},
\end{equation}
with $a_0$ being an integration constant. This geometry describes a
homogeneous space-time whose Riemann tensor has vanishing covariant
derivative, and thus all its higher derivative curvature scalars are
zero. The Ricci scalar is found to be $\frac{2}{\beta^2}$ while all
other curvature scalars are monomials of $\frac{2}{\beta^2}$ or
zero. The classical curvature singularity at $r=0$ is thus replaced
by an innocuous coordinate singularity, while the mass and the
electric charge are merged into the constant $\beta$ uniquely describing the curvature of the emerging
semiclassical geometry.

\subsection{The two dimensional subalgebra $(Q_5,Q_6)$}
According to the wave function \eqref{sol4} we have
\begin{equation}\label{T}
S(a,b,f)=\frac{2a^4b^2+2b^2f^4+2a^2b^2(1-2f^2)-4Qbf+2Q^2}{b(a^2-f^2)},\hspace{5mm}\Omega(a,b,f)=\frac{\lambda}{\sqrt{b(f^2-a^2)}},
\end{equation}
in which we have used again the numerical values \eqref{kapfix} for the
$\kappa_I$'s. A simple calculation based on the relation (\ref{D})
shows that ${\cal Q}=0$, i.e. the quantum potential vanishes in this
case as well. Also, the system of equation (\ref{G}) takes the form
\begin{eqnarray}\label{U}
\left\{
\begin{array}{ll}
\dfrac{8}{N}abb'=\dfrac{4a\left[a^4b^2-2a^2b^2f^2+b^2f^2(f^2-1)+2Qbf-Q^2\right]}{b(a^2-f^2)^2},\\\\
\dfrac{8}{N}\left(aba'+a^2b'\right)=\dfrac{2\left[a^4b^2+b^2f^4+a^2b^2(1-2f^2)-Q^2\right]}{b^2(a^2-f^2)},\\\\
\dfrac{8}{N}b^2f'=\dfrac{4\left[-a^4b^2f+a^2b(bf+2bf^3-Q)+f(-b^2f^4-Qbf+Q^2)\right]}{b(a^2-f^2)^2}.
\end{array}
\right.
\end{eqnarray}
Because of the vanishing quantum potential, we expect that the classical solutions satisfy the above equations. Indeed, a
combination of the first and third equations of the above system gives
\begin{equation}\label{V}
(bf)'=\frac{bf-Q}{b\, (a^2-f^2)},
\end{equation}
in which we have chosen again the gauge $N=2$. If, for the moment, we assume $(bf)'=0$,
the above equation yields $bf=Q$. This condition is
satisfied by the classical solutions
\begin{equation}\label{X}
b(r)=r,\hspace{5mm}f(r)=\frac{Q}{r},
\end{equation}
whereby, using them in the second equation of (\ref{U}), we get
\begin{equation}\label{Y}
2raa'+a^2=1-\frac{Q^2}{r^2},
\end{equation}
with the following solution for $a(r)$:
\begin{equation}\label{Z}
a(r)=\left(1+\frac{A}{r}+\frac{Q^2}{r^2}\right)^{1/2}.
\end{equation}
It is seen that after identifying the integration constant $A$ with
the mass parameter as $A=-2\, m$, we obtain the standard form of the
RN black hole line element \eqref{rnmet}.


\section{Explanation of the vanishing of the quantum potential} \label{sect5}
As it has become evident in the previous section, the quantum potential ${\cal Q}$ is different from zero only
in the case where $\widehat{Q}_3$ and $\widehat{Q}_4$ are imposed as ``simultaneous" eigenoperators.
In all other cases, the quantum potential becomes zero. Since this vanishing can be considered as a proof for a
kind of consistency (since the semiclassical solutions coincide with the classical ones), we are going,
in this section, to give an algebraic
explanation for it.

Let us start with the eigenvalue problem
\begin{align} \nonumber
& \widehat{Q}_I \Psi = \kappa_I\, \Psi \Rightarrow \widehat{Q}_I \left(\Omega\, e^{\ima S}\right) =\kappa_I\, \Omega\, e^{\ima S} \Rightarrow \\ \label{lin}
& \widehat{Q}_I \Omega + \ima\,\Omega\,  \widehat{Q}_I S = \kappa_I\, \Omega.
\end{align}
Due to the form of $Q_I$ \eqref{qlin}, \eqref{lin} can be split into
a real and an imaginary part, \be \ima\,  \widehat{Q}_I S =
\kappa_I\, \ee and \be \label{linima} \widehat{Q}_I\Omega =0, \ee
respectively.

The quantum potential is just \be {\cal Q} =\frac{1}{\Omega}\,
\square \Omega=\frac{1}{\Omega}\,\widehat{Q}_c \Omega
=\frac{1}{\Omega}
\left(\widehat{Q}_3^2+\widehat{Q}_2\widehat{Q}_5\right)\Omega, \ee
where the last equation holds due to \eqref{hamcasrel},
\eqref{hamoperator} (which are a consequence of both the constant
potential parametrization and the measure which allows the linear
operators to have the derivatives on the far right). Thus, the first
case of the Abelian 3d subalgebra is clear: the Laplacian is zero
because \eqref{linima} holds for each and every element of the
algebra ($I=2,3,5$), leading to a vanishing ${\cal Q}$.

For the 2d subalgebras:
\begin{enumerate}
\item ($\widehat{Q}_1$, $\widehat{Q}_2$) It must hold that
        \be
        \widehat{Q}_1\Omega =0\quad \text{and}\quad \widehat{Q}_2\Omega =0.
        \ee
        Thus, the quantum potential ${\cal Q}$ becomes (since $\widehat{Q}_2\widehat{Q}_5=\widehat{Q}_5\widehat{Q}_2$)
        \be
        {\cal Q} =\frac{1}{\Omega}\,\widehat{Q}_3^2 \Omega.
        \ee
        By definition \eqref{qlin} and the choice of measure ($\mu=\sqrt{\bbar{G}}$), the $\widehat{Q}_I$'s have all derivations on the far right. Moreover, by virtue of \eqref{intofmo}, we can see that $Q_3$ can be written as a linear combination (with functions) of $Q_1$ and
        $Q_2$, therefore dictating
        \be
        \widehat{Q}_3 = f\, \widehat{Q}_2 + \frac{1}{2\, b}\, \widehat{Q}_1.
        \ee
        The latter relation means that also $\widehat{Q}_3 \Omega =0$ and, as a result, again ${\cal Q}=0$.

\item ($\widehat{Q}_2$, $\widehat{Q}_3$) This case is straightforward:
        By assumption
        \be
        \widehat{Q}_2\Omega =0\quad \text{and}\quad \widehat{Q}_3\Omega =0,
        \ee which implies
        $\widehat{Q}_c \Omega=0$, thereby securing the vanishing of ${\cal Q}$.
\item ($\widehat{Q}_2$, $\widehat{Q}_5$) In this case,  one is left with
    ${\cal Q}=\frac{1}{\Omega}\, \widehat{Q}_3^2 \Omega$ and, apparently, a linear combination cannot
    be used (i.e. $Q_3 \neq A(q) Q_2+B(q) Q_5$). Nevertheless, the situation can be resolved by invoking
    the existence of the second Casimir invariant $\widetilde{Q}_C$ (equation \eqref{secondcas}) of the six
    dimensional algebra (which, thankfully, is identically zero in the differential representation corresponding
    to \eqref{intofmo}, otherwise there would be two quadratic constraints):

    Equation \eqref{linima} holds for $I=2$ and $I=5$, i.e.
    \be \label{q25}
    \widehat{Q}_2\Omega = 0 \quad \text{and} \quad \widehat{Q}_5 \Omega =0,
    \ee
    additionally \eqref{secondcas} can, demanding hermiticity and bearing in mind
    that $[\widehat{Q}_3,\widehat{Q}_4]=0$, be written in operator form as
    \be
    \widehat{Q}_2\widehat{Q}_6 +\widehat{Q}_6\widehat{Q}_2+ \frac{1}{2}(\widehat{Q}_1\widehat{Q}_5+\widehat{Q}_5 \widehat{Q}_1)- 2\,\widehat{Q}_4\widehat{Q}_3\equiv 0,
    \ee
    which, acting upon $\Omega$ yields (by virtue of \eqref{q25})
    \be \label{cas2a}
    \widehat{Q}_2\widehat{Q}_6\Omega+ \frac{1}{2}\widehat{Q}_5 \widehat{Q}_1 \Omega - 2\,\widehat{Q}_4\widehat{Q}_3\Omega=0.
    \ee
     Due to the algebra satisfied by the $Q_I$'s (in particular $[\widehat{Q}_2,\widehat{Q}_6]=\widehat{Q}_3$, $[\widehat{Q}_1,\widehat{Q}_5]=2\widehat{Q}_3$) one can bring $\widehat{Q}_2$ and $\widehat{Q}_5$ to the far right and thus \eqref{cas2a} reduces to
    \be \label{q41}
    \widehat{Q}_4\widehat{Q}_3\Omega=0.
    \ee
    At this stage, it is easy to check that
    \be \label{linq4}
    \widehat{Q}_4 =(f^2-a^2)\, b\, \widehat{Q}_2 +b\, \widehat{Q}_5,
    \ee
    which means that also
    \be \label{q42}
    \widehat{Q}_4\Omega=0.
    \ee
     Thus, relations \eqref{q41} and \eqref{q42} imply that $\widehat{Q}_3\Omega=\ima \,\lambda\, \Omega$, with $\lambda \in \mathbb{R}$ since $\widehat{Q}_3\Omega$ is imaginary and $\Omega$ is real.

    Let us now see what is the action of $\widehat{Q}_3$ on the full wave function
    $\Psi$:
    \be \label{q3psi}
    \widehat{Q}_3\Psi = \ima\,\lambda \Psi+ \ima \Psi \widehat{Q}_3 S.
    \ee
    We also calculate (using \eqref{q3psi})
    \be \label{q3sq}
    \widehat{Q}_3^2 \Psi= -\lambda^2 \Psi -2\, \lambda \Psi \widehat{Q}_3S - \Psi (\widehat{Q}_3S)^2+\ima \widehat{Q}_3^2 S.
    \ee
    The quadratic constraint on the wave function is
    \be \label{quantcon}
    \widehat{Q}_3^2\Psi + \widehat{Q}_2\widehat{Q}_5 \Psi - 4\Psi
    =0,
    \ee
    (the order of $\widehat{Q}_2$, $\widehat{Q}_5$ is irrelevant since they commute).
    By substitution of \eqref{q3sq} into \eqref{quantcon} we get
    \be \label{wdw1}
    -2 \lambda \Psi \widehat{Q}_3S - \Psi (\widehat{Q}_3S)^2+\ima\, \widehat{Q}_3^2 S +(\kappa_2\kappa_5-4-\lambda^2) \Psi=0.
    \ee
    If we break \eqref{wdw1} into real and imaginary part, we get
    \begin{align} \label{rewdw}
    & (\widehat{Q}_3S)^2+2\,\lambda\, \widehat{Q}_3S+\lambda^2+4-\kappa_2\kappa_5=0 \quad \text{and}\\ \label{imawdw}
    & \widehat{Q}_3^2 S =0,
    \end{align}
    respectively. Equation \eqref{rewdw} indicates that $\widehat{Q}_3 S$ is a constant and therefore \eqref{imawdw} is satisfied identically. The trinomial \eqref{rewdw} has the solution
    \be
    \widehat{Q}_3S=-\lambda\pm \ima\, \sqrt{4-\kappa_2\kappa_5}.
    \ee
    Under this, equation \eqref{q3psi} becomes
    \be
    \widehat{Q}_3\Psi = \pm \sqrt{4-\kappa_2\kappa_5} \Psi =\kappa_3\, \Psi.
    \ee
    So, $\Psi$ is an eigenfunction of $\widehat{Q}_3$ and \eqref{linima} must hold also for $I=3$, implying that $\widehat{Q}_3\Omega=0$ and therefore ${\cal Q}=0$.
\item ($\widehat{Q}_3$, $\widehat{Q}_5$) This is an easy case, since ${\cal Q}$ becomes zero immediately by $\widehat{Q}_3\Omega=\widehat{Q}_5\Omega=0$.
\item ($\widehat{Q}_5$, $\widehat{Q}_6$) Here $\widehat{Q}_5\Omega=\widehat{Q}_6\Omega=0$, which means that ${\cal Q}= \frac{1}{\Omega} \widehat{Q}_3^2\Omega$. But, $Q_3$ can be written as
    \be
    \widehat{Q}_3 = \frac{f}{a^2-f^2}\, \widehat{Q}_5 +\frac{1}{b\, (f^2-a^2)} \widehat{Q}_6,
    \ee
    which leads to $\widehat{Q}_3\Omega=0$ and, consequently, to ${\cal Q}=0$.
\end{enumerate}

\section{Discussion}

We have investigated the classical and quantum aspects of a
reparametrization invariant minisuperspace action which describes
the coupled Einstein-Maxwell system under the assumption of
spherical symmetry. At the classical level, the independent
dynamical variable is the radial coordinate $r$ while the two
unknown functions $a(r)$, $b(r)$ appearing in the general
spherically symmetric line element, span, along with the
electromagnetic potential variable $A_\mu=(f(r),0,0,0)$, the
configuration space of the (in principle) dynamical dependent
variables. The way the $r$-lapse function $n(r)$ enters the
Lagrangian \eqref{Lagv} and the line element \eqref{metric} makes manifest
the invariance of the action under arbitrary parametrizations
$r=h(\tilde{r})$. One can thus be led to the unique lapse
parametrization $n(r)=\frac{N(r)}{2\, a}$ in which the potential
$V(q)$ becomes constant, see \eqref{Lag}. The corresponding
supermetric \eqref{supmet} describes a Minkowskian configuration
space manifold and admits the six Killing vector fields
\eqref{Killing}. With their help we can, in the appropriate
phase-space, define the conditional symmetries \eqref{intofmo} which
have a vanishing Poisson bracket with the Hamiltonian
\eqref{hamiltonian} and are thus constant on the constraint surface
$\mathcal{H}\approx 0$ \eqref{autint}. The existence of the
homothetic vector \eqref{rnhom} provides us with another rheonomic
integral of motion \eqref{rehint}. It is noteworthy and interesting
that their counterparts in the velocity phase-space completely
describe the classical solution space as well as the two quadratic
relations \eqref{condcon} and \eqref{condcas2} corresponding to the
two existing Casimir invariants \eqref{casimir} and
\eqref{secondcas} of the algebra spanned by the six $Q_I$'s. Indeed,
using \eqref{rnsol} and the consistency relation $N= \frac{d}{dr}
\int\!\! N dr$, we algebraically (i.e. without ever solving the
corresponding differential equations) acquire the classical Reissner
- Nordstr\"om solution \eqref{rnmetb}, the quadratic relations
\eqref{condcon}, \eqref{condcas2} and the reparametrization
invariance since $b(r)$ remains undefined. Thus, we have the
solutions of the Einstein - Maxwell equations purely in terms of the
symmetries of the corresponding minisuperspace action.

At the quantum level, we demand Hermiticity under the unique natural
measure $\mu=\sqrt{\bbar{G}}$ in order to turn the conditional
symmetries $Q_I$ and the Hamiltonian constraint $\mathcal{H}$ into
operators \eqref{qlin}, \eqref{hamoperator}. In order to determine
which of the linear operators can be considered, we use the
integrability condition \eqref{intcon} which implies that only the
elements of certain subalgebras can be simultaneously applied on the
wave function $\Psi(a,b,f)$. We thus arrive at four distinct
families of quantum states (see the corresponding subsections of
section \ref{sect3}). Due to the well known problems of
interpretation of the wave function, we turn, in section
\ref{sect4}, to the semiclassical approximation in order to get a
glimpse at the fate of the classical singularity. We thus arrive at
the conclusion that the semiclassical equations of motion
corresponding to the asymptotic limit of the wave function
\eqref{S1} (derived from the subalgebra $\widehat{Q}_3$,
$\widehat{Q}_4$) indicate avoidance of the curvature singularity at
$r=0$ and that a horizon-like singularity does not appear (for
similar results in the context of Loop Quantum Cosmology see
\cite{Loop}). A very interesting occurrence is the vanishing of the
quantum potential $\cal{Q}$ in the other three cases, a fact that
leads to the semiclassical equations of motion giving rise to the
classical solution space. On the one hand, this is a negative
feature since it prohibits us from gaining some quantum information
at the semiclassical level; on the other hand, it can also be
considered as showing the consistency of the quantum theory in
consideration, and thus as a positive occurrence. It is thus
interesting to examine the reason for this vanishing of the quantum
potential. This is done in section \ref{sect5}: The main reason is
that the form of the wave function $\Psi=\Omega\, e^{\ima S}$
dictates that whenever a first order linear operator is applied as
$\widehat{Q} \Psi = \kappa \, \Psi$, the condition on $\Omega$ is
homogeneous, $\widehat{Q} \Omega=0$. This, in conjunction with the
two Casimir invariants and the particular form of the operators
fully explain the vanishing of the quantum potential $\cal{Q}$.

\end{document}